\documentclass[useAMS,usenatbib]{mn2e}

\usepackage {graphicx}
\usepackage {times}
\usepackage{pdflscape}
\usepackage{rotating}
\usepackage{natbib}
\usepackage{color}

\newcommand{\ltsima}{$\; \buildrel < \over \sim \;$}
\newcommand{\lsim}{\lower.5ex\hbox{\ltsima}}
\newcommand{\gtsima}{$\; \buildrel > \over \sim \;$}
\newcommand{\gsim}{\lower.5ex\hbox{\gtsima}}
\newcommand{\eg}{{\it e.g.,\ }}

\newcommand{\LCDM}{$\Lambda$CDM}

\newcommand{\nside}{{\rm n_{side}}}

\def\gtrsim{\mathrel{\hbox{\rlap{\hbox{\lower4pt\hbox{$\sim$}}}\hbox{$>$}}}}
\let\ga=\gtrsim
\def\lesssim{\mathrel{\hbox{\rlap{\hbox{\lower4pt\hbox{$\sim$}}}\hbox{$<$}}}}
\let\la=\lesssim

\newcommand{\nc}{\newcommand}

\nc{\be}[1]{\begin{equation}\mbox{$\label{#1}$}}
\nc{\bea}[1]{\begin{eqnarray} \mbox{$\label{#1}$}}
\nc{\Section}[2]{\section{#2}\label{#1}}
\nc{\Bibitem}[1]{\bibitem{#1}}
\nc{\Label}[1]{\label{#1}}

\nc{\vev}[1]{\langle #1 \rangle}

\nc{\eea}{\end{eqnarray}}
\nc{\ee}{\end{equation}}
\nc{\eeq}{\end{equation}}

\nc\map{{\sl WMAP\ }}

\def\simlt{\lower.5ex\hbox{\ltsima}}
\def\simgt{\lower.5ex\hbox{\gtsima}}

\def\likelihood{97~}
\def\likelihoodLCDM{9.9~}
\def\dchisqs{9.2}
\def\deltachisqthree{11.6, 16.1 and 20.7~}
\def\amplitudes{ 2.2 } 
\def\lamplitudes{ 3.4$\pm$1.1, i.e., 3.1$\sigma$ detection. }
\def\a_sigma{ 3.1}

\def\ltensigma{ 1.3}

\def\ltwelvesigma{ 1.7}

\title[Cross-correlation of WMAP and WISE]{Cross-correlation of WISE Galaxies with the Cosmic Microwave Background\footnotemark[0]
}
\author[Goto]{Tomotsugu Goto$^{1}$ 
,
 {Istv\'an Szapudi}$^{1}$,
 {Benjamin R. Granett}$^{2}$
\\
$^{1}$Institute for Astronomy, University of Hawaii
2680 Woodlawn Drive, Honolulu, HI, 96822, USA\\
$^{2}$Istituto Nazionale di Astrofisica - Osservatorio Astronomico di Brera, Via E. Bianchi 46, 23807 Merate, Italy}

\begin{document}

\def\Hg{H$\gamma$}
\def\Hd{H$\delta$}
\date{\today; in original form 2011 July 14}
\pagerange{\pageref{firstpage}--\pageref{lastpage}} \pubyear{2011}
\maketitle

\label{firstpage}

\begin{abstract}
We estimated the cross-power spectra of a galaxy sample from the {\sl
  Wide-field Infrared Survey Explorer} (WISE) survey with the 7-year
{\sl Wilkinson Microwave Anisotropy Probe} (WMAP) temperature
anisotropy maps.  A conservatively-selected galaxy sample
covers $\sim$13000$\sq\degr$, 
 with a median redshift of z=0.15.
Cross-power spectra show correlations between the two data sets with
no discernible dependence on the WMAP Q, V and W frequency bands.  We
interpret these results in terms of the the Integrated Sachs-Wolfe
(ISW) effect: for the $|b|>20\degr$ sample at $\ell=$6-87, we measure
the amplitude (normalized to be 1 for vanilla $\Lambda$CDM expectation)
of the signal to be  \lamplitudes We discuss other
possibilities, but at face value, the detection of the linear ISW
effect in a flat universe is caused by large scale decaying
potentials, a sign of accelerated expansion driven by Dark Energy.
\end{abstract}

\begin{keywords}
 cosmology:early universe
\end{keywords}

\section{Introduction}

Dark Energy (DE) is the principal puzzle of twenty-first century physics.
When identified with vacuum energy,  the quantum field theory prediction for it
is either 122 orders of magnitude too high or it is zero \citep{2009ARNPS..59..397C}. Both predictions
glaringly contradict astronomical observations suggesting that the amount
of DE today is of the same order of magnitude as that of Dark Matter. 
 The existence  of DE
rests on measurements of the cosmic microwave background fluctuations
(CMB), particularly by the {\sl Wilkinson Microwave Anisotropy Probe} 
satellite (WMAP) constraining essentially the total energy density
in the universe \citep{2011ApJS..192...18K}, and the low redshift acceleration of the universe
detected, e.g., through supernovae \citep{1998AJ....116.1009R}.
The evidence for DE appears to be overwhelming today, albeit indirect.
Most present and future observations
rely on standard candles and/or rulers to quantify the geometry and
expansion history, or growth history of the universe  (with the possible
exception of gravitational lensing).

The Integrated Sachs-Wolfe \citep[ISW][]{SachsWolfe,ReesSciama}
effect promises a unique avenue to
directly detect the effect of DE on CMB photons: as photons cross
potential wells, they suffer both blue and redshift cancelling each other
in a flat, DE-free universe where the potential wells are frozen.
However, when DE is present,
potential wells decay during the photon crossing, 
therefore the redshift fails to fully compensate
for the blueshift. The net result is a slight ``kick'' the photon receives.
This will  ultimately result in a tiny correlation of hot spots in the
CMB with large-scale structure, an effect which is orders of magnitude smaller than
the CMB correlations \citep{CrittendenTurok1996,PeirisSpergel2000}.

Such correlations were measured in the literature between the WMAP and the Sloan Digital
Sky Survey (SDSS) Luminous Red Galaxies
  \citep{ScrantonEtal2003,FosalbaEtal2003,Pad,GranettEtal2008a,PapaiEtal2011},
 APM galaxies \citep{FosalbaGaztanaga2004}, infrared galaxies 
\citep{AfshordiEtal2004}, radio galaxies \citep{NoltaEtal2004,Racc},
and the hard X-ray background 
\citep{BoughnCrittenden2004a,BoughnCrittenden2004b}. In addition to
the ISW effect, CMB photons become correlated with the LSS at high $\ell$
(small scales) due to the Sunyaev-Zeldovich effect, and lensing. The
latter was recently detected by \cite{SmithEtal2007}.

This analysis uses the first data release of the{\sl Wide-field
  Infrared Survey Explorer} \citep[WISE;][]{2010AJ....140.1868W}.  We use a sample of galaxies over
a survey area of $\sim$13000 $\sq\degr$ with a median redshift of z=0.15.
The combination of area and volume at moderate redshift has the
potential to detect (or reject) the ISW effect more clearly than
previous studies.

\section{Analysis and Results}
\subsection{Data}\label{Data}

\begin{figure}
\begin{center}
\includegraphics[scale=0.7]{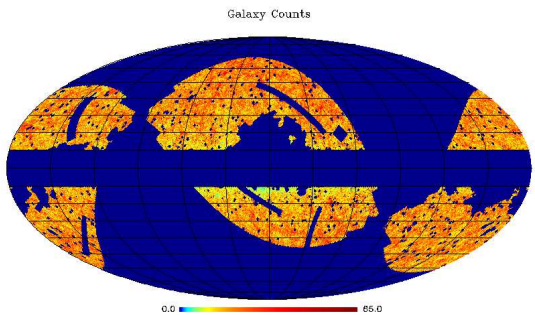}
\end{center}
\caption{
Galaxy counts from the WISE galaxy sample. Regions with zero galaxy
 counts are masked out by the object masks used for the WISE galaxy and
 WMAP sample. These are regions with $|b|<10$ deg, missing data, and star clusters.
}\label{fig:mask}
\end{figure}

\begin{figure*}
\begin{center}
\includegraphics[scale=0.3]{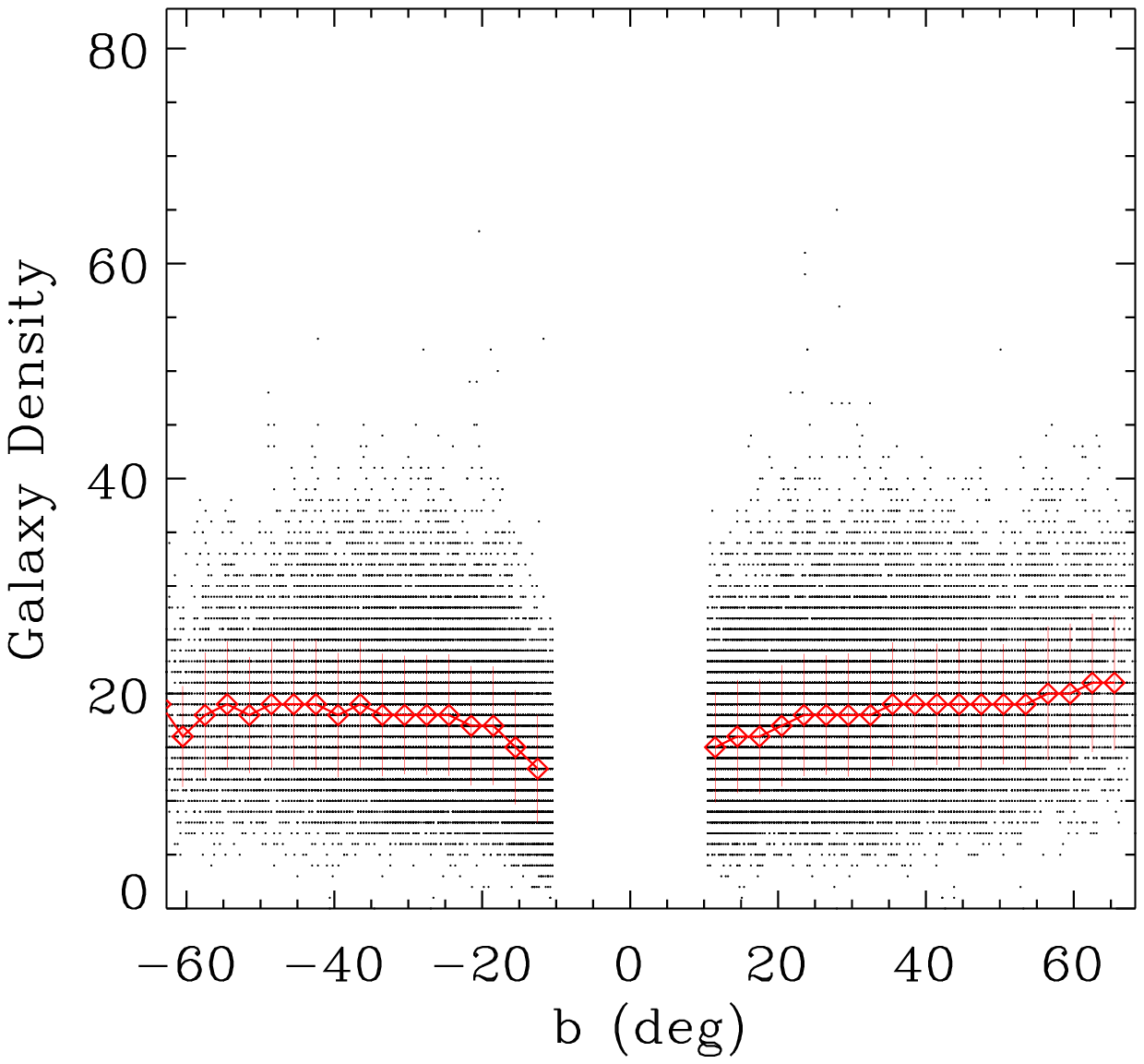}
\includegraphics[scale=0.3]{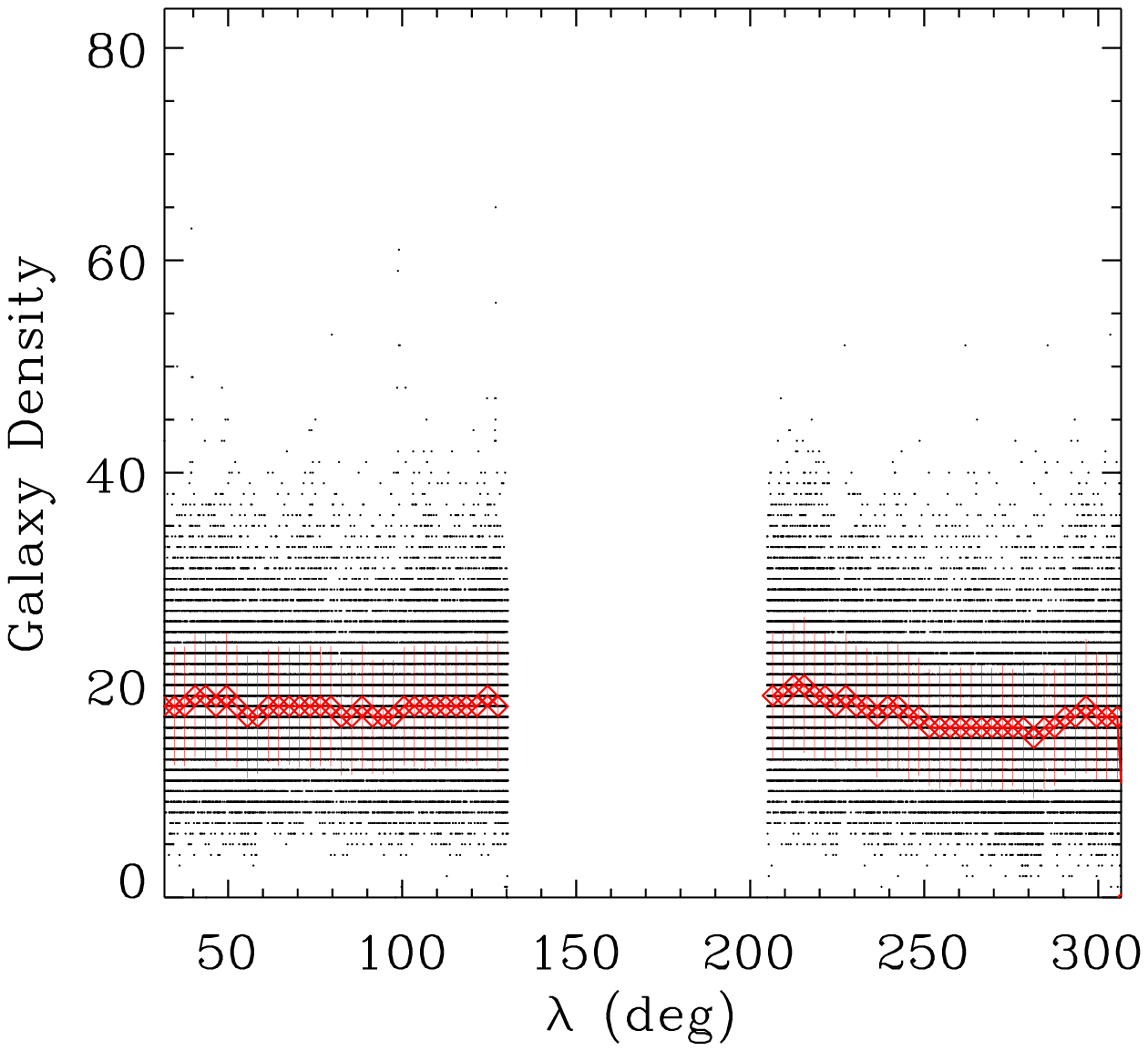}
\includegraphics[scale=0.3]{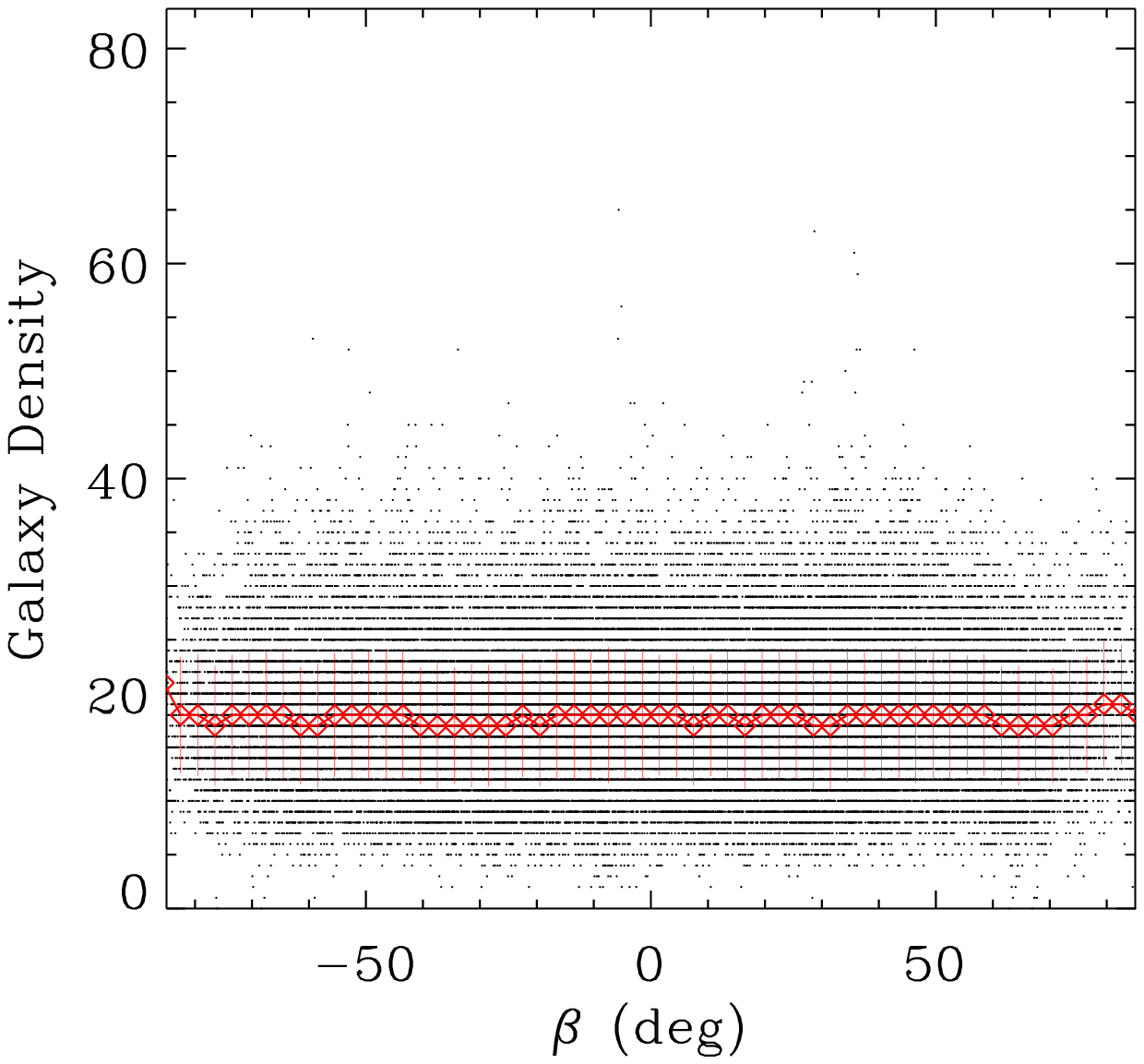}
\end{center}
\caption{ Galaxy number density (in 27\arcmin pixels) as a function of
  Galactic latitude $b$ and ecliptic coordinates $\beta$ and
  $\lambda$.  The red line connects median values, with RMS in the red
  vertical lines.  }\label{fig:density}
\end{figure*}

\begin{figure}
\begin{center}
\includegraphics[scale=0.25]{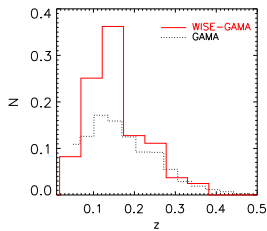}
\end{center}
\caption{
The normalized redshift distributions of the WISE galaxy sample 
(red solid line) and the GAMA spectroscopic sample (black dotted line).
}\label{fig:zhist}
\end{figure}

For the CMB, we use the WMAP 7-year data set
\citep{2011ApJS..192...18K}.  We use the Q, V and W foreground removed
maps.  We rebin the maps in Healpix format to $\nside=128$ for a pixel
size of 27\arcmin/pixel \citep{Healpix}.  We apply the Temperature Analysis Mask
provided by the WMAP team to exclude regions of Galactic emission and known point
sources leaving about 78\% of sky.

The WISE satellite surveys
the entire sky at wavelengths of 3.4, 4.6, 12, and 22 $\mu$m ($W1$
through $W4$), with
the point source sensitivities of 0.08, 0.11, 1, and 6 mJy, respectively. 
We use the first data release
issued on April 14, 2011, containing about half the sky.

We need to remove stars from the WISE object catalogue.  We find that
most stars ($|b|<10\degr$) have a $3.4-4.6\mu m$ color less than 0.2.
Therefore, we apply a color cut of $3.4-4.6\mu m>$0.2 to select
galaxies.  We further apply a color selection of $4.6-12\mu m>$2.9,
which removes a greater number of stars at the expense of a small number of
galaxies.  Due to the sun-synchronous orbit of the satellite, WISE
data are deeper in pole regions. To select a uniform galaxy sample, we
only use galaxies with $W1_{3.6\mu m}<15.2$ mag.

We construct Healpix maps of the galaxy counts matching our CMB maps, shown on
Fig. \ref{fig:mask}.  We apply the WMAP mask and additionally mask out
regions with missing data and unusually high object-counts such as
star clusters and the Magellanic clouds.  These can be seen as thin
stripes in Fig. \ref{fig:mask}, totaling 613 $\sq\degr$ of area.  We also
mask cells with more than 10 flagged sources (as persistent, halo,
ghost, or spike).  This helps to exclude cells contaminated by bright
nearby galaxies and stars, as visual inspection of pixels in the
tail confirmed.  We do not use regions with $|b|<10\degr$ due
to strong stellar contamination.  Our galaxy catalog contains 1.2
million galaxies over an area of 13622 $\sq\degr$ and the median galaxy
density is 86 deg$^{-2}$.

We investigate gradients in the galaxy
density in Fig. \ref{fig:density} as a function of Galactic latitude
$b$, and ecliptic coordinates $\beta$ and $\lambda$.  
There appears to be a significant decrease in the galaxy number 
density at $10<|b|<15$, reaching 30\% at $|b|=10\degr$.  
At this low galactic latitude,  Milky Way stars
can mask background galaxies and confuse photometry.  This effect has been studied in optical bands \citep{Ross2011}, but it is more severe in WISE due to the broader point spread function.
In the middle panel, the galactic centre is at  $\lambda\sim$270$\degr$, and the same effect can be seen.
To test how this might affect our results, 
we analyse maps with three different galactic cuts ($|b|>$20\degr, 15\degr
and 10\degr) in Section \ref{sec:ISW}.
  The WISE data become deeper toward the pole regions, but there is no
significant change in galaxy density as a function of $\beta$ (right
panel).  

To estimate the expected ISW signal, we need to know the redshift
distribution of our galaxy sample, and the galaxy bias.  We have
cross-matched the WISE galaxy sample with GAMA \citep[Galaxy and Mass
  Assembly;][]{driver11} sample to estimate a preliminary redshift
distribution.  GAMA is a spectroscopic sample of 205,000 galaxies
observed with AA$\Omega$ 4m telescope. The survey depth is as deep as
$r_{AB}<20.5$ mag, and thus, deep enough to estimate the redshift
distribution of the WISE galaxy sample.  
In the overlapping regions, 83\% of the WISE galaxy sample had spectroscopic counterpart
in GAMA. 
Fig. \ref{fig:zhist} shows the
redshift distribution of the WISE galaxy sample with GAMA
counterparts. The dotted black histogram shows the redshift
distribution of the GAMA sample, which has a much longer tail toward
higher redshift.  The median redshift of the WISE-GAMA sample is
z=0.148.

\subsection{Galaxy bias}\label{sec:bias}
\begin{figure}
\begin{center}
\includegraphics[scale=0.4]{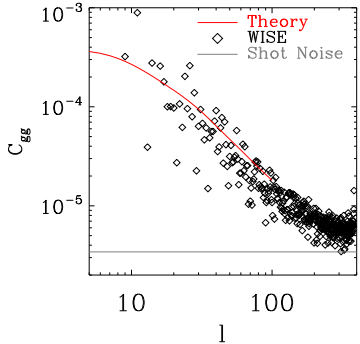}
\end{center}
\caption{
Auto correlation of the WISE galaxy sample ($|b|>20\degr$). The red solid curve is a
fit of a theoretical matter power spectrum from CAMB with the latest
cosmological parameters for $\Lambda$CDM.
The gray solid line is a Poisson shot noise.
}\label{fig:bias}
\end{figure}

In order to determine galaxy bias, we measure the galaxy-galaxy
angular power spectrum.  The form of the angular power spectrum can be
theoretically computed given cosmological parameters.  We use the WMAP
7-year $\Lambda$CDM cosmological parameters ($H_0=70.4$ km/s/Mpc,
$\Omega_m=0.272$).  However, in linear theory, bias cannot be
determined independently of $\sigma_8$, 
the power spectrum is normalized as $C_l\propto(b\sigma_8)^2$, 
where $b$ relates galaxy and matter overdensities as $\delta_g=b\delta_m$.
We therefore use a fixed
value of $\sigma_8$=0.80, and fit the galaxy-galaxy angular
correlation only for bias.  We use CAMB with Halofit \citep{LewisEtal2000,Smith03} to
generate non-linear matter powerspectra for our cosmological model,
and obtain three-dimensional spectra at the median redshift of
z=0.148.  The angular power spectrum is a projection of the
three-dimensional power spectrum with a kernel that depends on the
redshift distribution:
\begin{equation}
C_l^{gg}=b_g^2 \frac{2}{\pi}
\int \left[\int r^2 \phi(r) j_l(kr) dr\right]^2 k^2 P(k) dk,
\end{equation}
with $\phi(r) \propto \frac{dN(r)}{dz}\frac{dz}{dV}$ for comoving
coordinate $r$ normalized such that $\int \phi(r)r^2dr=1$ and $j_l$ is
the spherical Bessel function.  We make use of the small-angle limit
with the Limber equation to evaluate the power spectrum at $l>20$, see
\eg\citet{AfshordiEtal2004}.

We compute the galaxy-galaxy angular power spectrum of the WISE galaxy
sample using SpICE
\citep[Spatially Inhomogeneous Correlator
  Estimator;][]{SzapudiEtal2001a,SzapudiEtal2001b}, a fast quadratic
estimator which inverts the coupling matrix of the $C_l$'s in pixel
space where it is diagonal.  
The resulting power spectrum of the WISE galaxy sample is shown in
Fig. \ref{fig:bias}.  
 The Poisson shot noise is given by $1/N$, where $N$ is the mean number
 of galaxies per steradian. In our sample, $1/N=$3.4$\times 10^{-6}$,
 which is much smaller than the correlation at $l=2-100$.
We subtract the Poisson noise (gray line), then fit the amplitude of the model template over
the range $l=2-100$. 
  We find a bias parameter of $b_g\equiv b \sigma_8/0.8 = 1.06 \pm0.05$.

\subsection{WISE-WMAP Cross Power Spectrum}\label{sec:ISW}

We cross-correlate the WISE galaxy sample with the foreground-removed Q, V, and W maps from WMAP.  
Because of the declining galaxy density at low galactic
lattitude $b$ (Fig.\ref{fig:density}), we show results in three different samples, $|b|>10$, $|b|>15$ and $|b|>20$.
But first, we show results with the $|b|>20$ sample.
The resulting angular power spectra are shown in Fig. \ref{fig:isw}. The spectrum (black
dots) band powers are binned 0.15 dex logarithmic bins.
The boundaries are $l$=6,8,11,16,22,31,44,61 and 87 such that the first 
band includes $l=$6,7, etc.
We do not use data at
$l\geq$87 since these small scales should not be sensitive to the ISW
effect. 

In order to estimate uncertainty in the WISE-WMAP cross power
spectrum, we have run 1000 simulations.  Using {\ttfamily synfast} and
the latest $\Lambda$CDM cosmological parameters, we create 1000 random
CMB maps. Cross-correlating these random CMB maps with the WISE
galaxy catalog, we estimated the covariance matrix. 
It is dominated by
diagonal terms, i.e.  neighboring band powers are uncorrelated to a 
good approximation.

The error bars plotted in Fig. \ref{fig:isw} are given by the diagonal
elements of the covariance matrix.  The large-scale modes at $l<6$ are
not well constrained on the cut sky.  The size of a contiguous survey region in the WISE map
(Fig. \ref{fig:mask}) is $\sim$60\degr, or 30\degr in radius.  This
corresponds to a limiting scale of $l\sim180\degr/\theta\sim$6. Therefore,
we do not have strong constrains at $l<6$. For simplicity we exclude 
the large-scale modes at $l<6$ in the analysis.

The $C_\ell$'s are consistent among different WMAP frequencies, while
most foregrounds from Milky Way emission should exhibit some color
dependence. 

Theoretical expectation for the ISW power spectrum from the the latest
cosmological parameters for $\Lambda$CDM are plotted with the red solid line.  
We adopted bias the parameter of $b_g=1.06$ estimated from auto-correlation in
Section \ref{sec:bias}.  The galaxy density-CMB cross power spectrum
is given by \citep{Cooray2001b},
\begin{eqnarray}
\lefteqn{C_l^{gT}=T_{CMB}\frac{3 \Omega_m H_0^2}{c^2} b_g \frac{2}{\pi} \int k^2 dk P_k \times} \nonumber\\
& &  \int k^{-2} \frac{d(1+z)D_1(z)}{dr} j_l(kr) dr \int r'^2 \phi(r') j_l(kr') dr'
\end{eqnarray}
where $D_1(z)$ is the linear growth factor.

The measured power spectrum is higher than the \LCDM~prediction.  To
quantify this we scale the model with a free amplitude parameter. 
In Fig. \ref{fig:isw}, the best $\chi^2$-fit was in the orange line, which is 
\amplitudes $\sigma$ larger than the original theoretical expectation.

\subsection{Significance tests}

 We now test the following three hypotheses for consistency with 
the data:
(a) theoretical best-fit ISW (orange line in Fig. \ref{fig:isw}),
(b) theoretical ISW from vanilla $\Lambda$CDM 
(red line) 
(c) Null hypothesis: no power, i.e. no ISW effect.

We follow \citet {francis2010} and compute $\chi^2$ values using the covariance matrix for the three hypotheses.  In Fig. \ref{fig:isw},
 we see a high power at $l$=9 and 12, that drops at $l=7$.
The significances of correlation at $l=9$ and 12 are 
\ltensigma ~and \ltwelvesigma $\sigma$, respectively. 
While the significance of an individual band power is affected by
the binning, the final $\chi^2$ values and likelihoods are robust.
To check for this, tested alternative band powers with boundaries at $l$=6,9,12,17,24,48,67 and 95 (purple squares in
Fig.\ref{fig:isw}); none of our numerical results or conculsions changed.

 The $\chi^2$ statistic is given by 
$\chi^2=({\bf x}-{\bf t}_i)^T{\bf C}^{-1}({\bf x}-{\bf t}_i)$
  for the observed data vector ${\bf x}$ and a given theoretical expectation  ${\bf t}_i$ for model $i$.
For  $l$=6-87 with 8 data points, the $\chi^2$ values are \deltachisqthree for (a), (b), and (c), respectively.
The  $\chi^2$ values decrease significantly for the ISW models (a) and (b) compared with the null hypothesis.
Computed  $\Delta \chi^2$ (Ratio of evidence) are 
\dchisqs
~for  $l$=6-87.%
 ~Generally, $\Delta\chi^2\ge$3 can be considered as strong evidence.
 Our best-fit measurement has $\Delta\chi^2>$3 in all cases. 

In terms of the likelihood ratio, $e^{1/2 \Delta\chi^2}$, we obtain \likelihood and \likelihoodLCDM as shown
in Table\ref{tab:chisq8}.  The likelihood
ratio is considered significant if it is greater than 5.
According to these ratios of evidence, the ISW hypotheses (a) and (b) are
preferred over the null hypothesis.

For the model (a), these numbers can be compared with the significance of
the amplitude fit itself.  
We computed 1-$\sigma$ errors of the fit of the theoretical
ISW model to the data, by estimating the 68 percentile of the $e^{-1/2
\chi^2}$ 
distribution.  We also checked that the measured errors are
consistent with the width of the best-fit Gaussian.  Results in Table
\ref{tab:amplitude} show that the best-fit model is \a_sigma
$\sigma$ from the null hypothesis.

\subsection{Systematic tests}

We have shown results with $|b|>20\degr$ sample. In this section, 
we explore if changing the galactic cut affects our results.

With a $|b|>15\degr$ cut, we have $\sim$1700 $\sq\degr$ more area than the
$|b|>20\degr$ cut.  The resulting amplitude of the best-fit ISW model is
3.2$\pm$1.1, i.e., 2.9 $\sigma$, in agreement with the result
from $|b|>20\degr$. As shown in Table \ref{tab:chisq8}, the $\Delta \chi^2$ and
likelihood ratio are decreased, but are not significantly changed. At $|b|\sim$15\degr the galaxy density starts
to decline towards the galactic plane, and thus, the correlation signals have
been reduced some, but overall, we obtained consistent results with the
$|b|>20\degr$ analysis.

Next we used $|b|>10\degr$ sample.  This sample adds another $\sim1600$
$\sq\degr$. As shown in Tables \ref{tab:chisq8} and \ref{tab:amplitude},
the signal becomes lower and the error is larger giving a worse
significance of 2.0$\sigma$.  This is most-likely due to the
systematic decrease in galaxy density at low $b$, where artefacts of
bright stars and high stellar density decrease the galaxy density.
Large-scale gradients in the galaxy distribution can increase the
measurement errors because they are amplified by cosmic variance of the CMB
on large angular scales.

To test this interpretation, we artificially flattened the galaxy
density as a function of $b$, by dividing the galaxy density by the mean
galaxy density at each $b$.  This process can be considered as a high
pass filtering that removes large-scale gradients. As shown in Tables
\ref{tab:chisq8} and \ref{tab:amplitude}, the significance is
increased to 3.2 $\sigma$, becoming consistent with values we measured
with $|b|>20\degr$ and 15$\degr$ samples.

 To the opposite direction, we tried  $|b|>25\degr$, again obtaining a consistent amplitude of  3.5$\pm$1.2 (Tables
\ref{tab:chisq8} and \ref{tab:amplitude}).
These tests show that our results are robust, 
obtaining $\sim$3 $\sigma$ significance regardless the
choice in the galactic cut $b$.

\begin{figure}
\begin{center}
\includegraphics[scale=0.6]{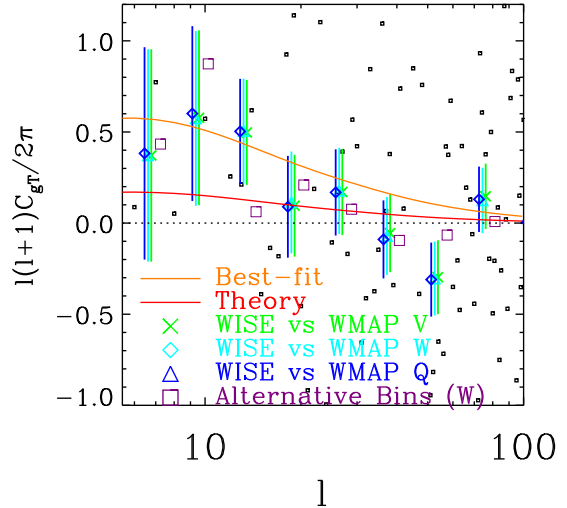}
\end{center}
\caption{
Power spectra between the WISE galaxy sample  ($|b|>20\degr$) and WMAP (V,Q,W)
maps. The data in original resolution (black dots) are binned in 0.15
dex logarithmic bins and shown with error bars. The red solid line is
 theoretical expectation with the bias of 1.06, while the orange line is
 the best fit to the observed data. The fit used 8 ($l$=6-87) data
 points. The purple line is from WMAP W data in different binning.
}\label{fig:isw}
\end{figure}

We briefly discuss stellar contamination next.
Stars are not clustered, but star counts might be correlated with the CMB due to galactic contamination. 
According to Fig.~\ref{fig:density}, the galaxy density is constant at 
$|b|>$20, although there is a slight negative gradient at $|b|<$20.
We argue that is likely to be due to artefacts and confusion with stars.  
The robustness of the results with respect to galactic cuts, however, implies 
that the effect on the correlations is negligible for the
foreground subtracted CMB maps we use. We do not 
see any significant colour dependence either between WMAP frequencies 
($V,W$, and $Q$) in Fig.~\ref{fig:isw}. 
These points suggest that the stellar contamination is small, 
or at least uniform across our survey area.  
In the worst case the contamination would be 17\%, if all unidentified
GAMA sources were stars, but it is likely to be much less.

\begin{table}
\begin{center}
\caption{Significances from $\chi^2$ tests at $l=6-87$.}
\label{tab:chisq8}
\begin{tabular}{ccccccc}
  \hline
  Galactic cut  && $\chi^2$  & d.o.f.&   $\Delta\chi^2$ &  Likelihood ratio \\ 
 \hline
 \hline
 $|b|>25\degr$ & Best-fit ISW          & 10.2 &7 & 8.7  & 77    \\
&ISW with $\Lambda$CDM & 14.7 &8 & 4.2  & 8.2 \\
&Null hypothesis       & 18.9 &8 & 0    & -\\
\hline
 $|b|>20\degr$ &Best-fit ISW          & 11.6 &7 & \dchisqs  & \likelihood    \\
&ISW with $\Lambda$CDM & 16.1 &8 & 4.6  & \likelihoodLCDM \\
&Null hypothesis       & 20.7 &8 & 0    & -\\
\hline
 $|b|>15\degr$ &Best-fit ISW          & 11.0 &7 & 8.1  & 59  \\
&ISW with $\Lambda$CDM & 14.9 &8 & 4.3  & 8.6 \\
&Null hypothesis       & 19.2 &8 & 0    & -\\
\hline
 $|b|>10\degr$ &Best-fit ISW          & 9.1  &7 & 3.8  &  6.6  \\
&ISW with $\Lambda$CDM & 11.1 &8 & 2.6  &  3.2 \\
&Null hypothesis       & 13.7 &8 & 0    & -\\
\hline
 $|b|>10\degr$ &Best-fit ISW          & 14.3 &7 & 10.2 &  166\ \\
with flattening&ISW with $\Lambda$CDM & 19.1 &8 & 5.4  &  14.8 \\
&Null hypothesis       & 24.5 &8 & 0    & -\\
 \hline
\end{tabular}
\end{center}
\end{table}

\begin{table}
\begin{center}
\caption{Amplitude and errors of the best-fit ISW model at $l$=6-87.}
\label{tab:amplitude}
\begin{tabular}{ccccccc}
  \hline
  Sample &   Area ($\sq\degr$) &Amplitude and error & $\sigma$\\
 \hline
$|b|>25\degr$& 8495 &   3.5$\pm$1.2 & 3.1 \\
$|b|>20\degr$& 10337 &   3.4$\pm$1.1 & 3.1 \\
$|b|>15\degr$& 12032 &  3.2$\pm$1.1 & 2.9 \\
$|b|>10\degr$&13622 &  2.6$\pm$1.3   & 2.0\\
$|b|>10\degr$ with flattening &13622  & 3.2$\pm$1.0 & 3.2\\
\hline
\end{tabular}
\end{center}
\end{table}

\section{Discussion and Conclusions}\label{discussion}
The significant correlations between WMAP and WISE on large scales may
have contributions from many sources including the Milky Way,
extragalactic point sources, even zodiacal light.  We have attempted
to minimize contamination by constructing a clean galaxy sample from
WISE.  We find no systematic trends in galaxy number density with
ecliptic latitude and longitude (Fig. \ref{fig:density}).  We use the
WMAP foreground reduced CMB maps to reduce the sensitivity to Milky
Way emission in WMAP bands.  The primary sources (dust, synchrotron
and free-free emission) have characteristic spectral shapes across the
WMAP frequency bands.  The excellent agreement between
the Q, V and W spectra (Fig. \ref{fig:isw}) leads us to believe that
Galactic foregrounds are unimportant. 

Extragalactic point sources below the WMAP detection limit
could also contaminate the signal. Since WISE is a near-IR
survey, galaxies detected in WISE may also emit strongly at WMAP
frequencies.  Such sources would produce a characteristic signal in
the power spectrum that could become important on small scales.  The
zeroth order estimate of this effect is a Poisson term with a 
flat $C_l$ spectrum; no
evidence for this is seen at $l<87$.  Similarly, the
Sunyaev-Zeldovich effect is expected to appear at higher $l$ and has
negligible effect at $l<87$ \citep{2004MNRAS.350L..37F,AfshordiEtal2004}.

We leave detailed discussion of the cosmological significance to a
future work, a few comments will suffice here.  Our best fit ISW model
is \amplitudes\ $\sigma$ higher than \LCDM\ expectation. This is
consistent with many previous cross-correlation measurements
\citep{Ho08,gian08,GranettEtal2009} that had higher amplitude than
vanilla \LCDM\ prediction.  
Possible reasons include
uncertainties in cosmological parameters or our galaxy bias model.
Additionally, the measurement is subject to significant cosmic variance
originating from both the CMB and LSS fields \citep{PapaiEtal2011}.
At present there is at most a mild tension between vanilla \LCDM~and 
measurements that is 
not fully understood.

In the \LCDM~model, the amplitude of the ISW signal is well
constrained by our knowledge of $\sigma_8$, the Hubble constant and
the matter density.  However, we do face uncertainty in the redshift
distribution and its effect on our estimate of the linear bias and the
amplitude of the ISW spectrum.  We obtained the $dN/dz$ by matching
the WISE galaxy sample to the GAMA spectroscopic sample.  Although the
GAMA sample is deep, it is an optically selected sample while WISE is
sensitive to the near IR.  Thus, GAMA may be missing a population of
WISE galaxies such as heavily obscured one.  We checked the fraction
of our WISE galaxy sample detected by GAMA was 83\%.  While 83\% is a
large fraction, the redshift distribution of the rest 17\% is unknown.

However, even if we assume this 17\% of galaxies are all at higher
redshift than 0.148, the median only changes to z=0.155.  More
generally, to test whether erroneous redshift distribution can affect
bias estimate, we have artificially shifted the median redshift from
z=0.09 to z=0.22, a value clearly beyond reasonable.  The change in
bias is within 40\%.  Therefore, the errors in the redshift
estimation alone can only account for a fraction of the shift
between the original and best-fit theoretical ISW estimations.
Furthermore, the amplitude of the ISW spectrum is relatively 
insensitive to the median redshift over this range.

The most relevant measurements in the literature to compare with are
from 2MASS for which weak and null detections of the ISW signal have
been claimed \citep{rassat07,francis2010}.  However, the median
redshift of 2MASS is z=0.07 and in terms of comoving volume, the WISE
sample is 4.6 times larger.  \citet{francis2010} selected a higher
redshift sample ($0.2<z<0.3$) using photometric redshifts but it is
also much more sparse.  The WISE sample may be situated in an ISW
`sweet spot': it is a large volume and samples a low redshift when
dark energy has come to dominate the cosmic expansion.  Indeed, the
WISE survey seems in better agreement with higher redshift datasets
\citep{Ho08,gian08}.  However, the \LCDM~ expectations are at the
detection limit for current surveys.  It is possible that we are only
measuring the ISW signal from ``lucky'' statistical fluctuations that
push the signal above the typical 2-$\sigma$ detection threshold
leading to biased conclusions if taken at face value.  Future surveys
that expand on the sky coverage and redshift range in the next decade
will provide a clearer picture on the integrated Sachs-Wolfe effect
and the role of dark energy.

We thank the anonymous referee for many insightful comments. 
We acknowledge financial support from the NASA grant NNX10AD53G and the Polanyi program of the Hungarian National Office for the Research and Technology (NKTH).

\bibliography{wise} 
\bibliographystyle{mnras}

\label{lastpage}

\end{document}